\documentclass[twocolumn,nofootinbib,amsmath,amssymb,a4paper]{revtex4}

\usepackage{graphicx}
\usepackage{dcolumn}
\usepackage{bm}

\begin{document}

\title{Quenched lattice calculation of the $B\rightarrow D \ell \nu$ decay rate}

\author{G.M. de Divitiis$^{a,b}$, E. Molinaro$^{c}$, R. Petronzio$^{a,b}$, N. Tantalo$^{b,d}$}
\affiliation{\vskip 10pt
$^{a}$~Universit\`a di Roma ``Tor Vergata'', I-00133 Rome, Italy\\
$^{b}$~INFN sezione di Roma ``Tor Vergata'', I-00133 Rome, Italy\\
$^{c}$~SISSA, I-34014 Trieste, Italy\\
$^{d}$~Centro Enrico Fermi, I-00184 Rome, Italy
}%

\begin{abstract}
We calculate, in the continuum limit of quenched lattice QCD, the form factor that enters in
the decay rate of the semileptonic decay $B\rightarrow D \ell \nu$. Making use of
the step scaling method (SSM), previously introduced to handle two scale problems
in lattice QCD, and of flavour twisted boundary conditions we extract $G^{B\rightarrow D}(w)$
at finite momentum transfer and at the physical values of the heavy quark masses.
Our results can be used in order to extract the CKM matrix element $V_{cb}$ by the
experimental decay rate without model dependent extrapolations.
\end{abstract}

\maketitle

\section{Introduction}
The central  goal of flavour physics is the determination of the 
Cabibbo--Kobayashi--Maskawa~\cite{Cabibbo:1963yz,Kobayashi:1973fv}
quark mixing matrix. A set of redundant and precise measurements can also provide informations about possible new physics~\cite{Bona:2006ah}. In turn, precise measurements need an adequate 
theoretical determination of hadron matrix elements of the weak currents.
Non perturbative tools and in particular lattice QCD can eventually provide the required precision.
In this letter we address the determination at finite momentum transfer of the form factor 
that enters in the semileptonic heavy meson decay rate. 
In the infinite quark mass limit it is parametrized by the Isgur-Wise 
function~\cite{Isgur:1989ed}. In particular, we calculate the matrix elements entering the 
determination of $V_{cb}$.
The calculation that we present  misses the effects of dynamical fermions and is not the final 
one. Nevertheless, it accomplishes for the first time the calculation at finite momentum 
transfer and at the physical values of the heavy quark masses, allowing to compare 
experimental data without additional model dependent extrapolations. 

We use a method~\cite{Guagnelli:2002jd} already applied successfully to the determination of heavy quark masses and decay 
constants~\cite{deDivitiis:2003wy,deDivitiis:2003iy,Guazzini:2006bn}, called the step scaling 
method (SSM), and special boundary conditions, called flavour 
twisted~\cite{deDivitiis:2004kq}, that shift by an arbitrary amount the discretized set of lattice momenta (see also~\cite{Sachrajda:2004mi,Flynn:2005in}). The step scaling method allows to reconcile large quark masses  with adequate lattice resolution and large physical volumes; the flavour twisted boundary conditions allow to perform a calculation at non zero momentum transfer with good accuracy.
We present the main results of our computation and stress their phenomenological implications; 
a more detailed discussion on technical aspects will be presented elsewhere~\cite{comingsoon}.

\section{Form factors and decay rate}
The semileptonic decay of a pseudoscalar meson into another pseudoscalar meson is mediated by
the vector part of the weak $V-A$ current. The corresponding matrix element can be 
parametrized in terms of two form factors. Among possible parametrizations we chose the 
following one
\begin{eqnarray}
\frac{\langle \mathcal{M}_f\vert \ V^\mu \ \vert \mathcal{M}_i\rangle}{\sqrt{M_i M_f}}=
(v_i+v_f)^\mu \ h_+ + (v_i-v_f)^\mu \ h_- \nonumber
\end{eqnarray}
where $M_{i,f}$ and $v_{i,f}=p_{i,f}/M_{i,f}$ are the meson masses and $4$-velocities.
The form factors depend upon the masses of the initial and final particles and upon
$w\equiv v_f\cdot v_i$
\begin{eqnarray}
&&h_\pm \equiv h_\pm^{i\rightarrow f}(w)\equiv h_\pm(w,M_i,M_f)
\nonumber \\ \nonumber \\
&&1\le w \le (M_i^2+M_f^2)/2M_iM_f \nonumber
\end{eqnarray}
In the case where $M_i$ is the $B$ meson mass and $M_f$ is the $D$ meson mass 
the maximum value of $w$ is around $1.6$.
In the infinite mass limit the form factors become
\begin{eqnarray}
&&h_+^{i\rightarrow f}(w)=\xi(w) 
\nonumber \\ 
&&\qquad \qquad \qquad \qquad \qquad M_f,M_i\rightarrow\infty
\nonumber \\ 
&&h_-^{i\rightarrow f}(w)=0 \nonumber
\end{eqnarray}
where $\xi(w)$ is the universal Isgur-Wise function; 
the conservation of the vector current implies $\xi(1)=1$. In what follows we 
will find deviations from such a limit that will allow a precise discussion on the onset of the HQET regime~\cite{comingsoon}.

The differential decay rate for the process $B\rightarrow D\ell\nu$, in the case of
massless leptons, is given by
\begin{eqnarray}
&&\frac{d\Gamma^{B\rightarrow D\ell\nu}}{dw}=
\nonumber \\ \nonumber \\  
&&\vert V_{cb}\vert^2 \frac{G_F^2}{48\pi^3}(M_{B}+M_{D})^2M_{D}^3(w^2-1)^{3/2}
\left[ G^{B\rightarrow D}(w)\right]^2
\nonumber
\end{eqnarray}
The phase space factor $(w^2-1)^{3/2}$ in the decay rate makes its 
experimental determination harder as $w$ approaches $1$ and the value at zero recoil is 
obtained from an extrapolation.
The function $G^{B\rightarrow D}(w)$, 
\begin{displaymath}
G^{i\rightarrow f}(w)=h^{i\rightarrow f}_+(w)\ -\
\frac{M_f-M_i}{M_f+M_i}\ h^{i\rightarrow f}_-(w)
\end{displaymath}
is needed in order to extract $V_{cb}$ by the measurement of the decay rate.
Previous lattice calculations by the Fermilab 
collaboration~\cite{Hashimoto:1999yp,Okamoto:2004xg} quote the value of 
$G^{B\rightarrow D}(w)$ only at zero recoil where it can be extracted with good statistical accuracy by using the so called "double ratio" technique. 
In the following we calculate $G^{B\rightarrow D}(w)$ in the range  $1\le w \le 1.2$ that 
includes values of $w$ where experimental data are available.

\section{Lattice Observables} \label{sec:notations}
We have carried out the calculation within the $O(a)$ improved Schr\"odinger Functional 
formalism~\cite{Luscher:1992an,Sint:1993un} with $T=2L$ and vanishing background fields. 
In order to fix the notations, we introduce the following lattice operators
\begin{eqnarray}
&&O_{rs}=\frac{a^6}{L^3}\sum_{{\bf y},{\bf z}}{\bar{\zeta}_r({\bf y})\gamma_5\zeta_s({\bf z})}
\nonumber \\
&&O_{rs}^\prime=\frac{a^6}{L^3}\sum_{{\bf y},{\bf z}}{\bar{\zeta}^\prime_r({\bf y})\gamma_5
\zeta_s^\prime({\bf z})}
\nonumber \\ \nonumber \\ \nonumber \\
&&A^0_{rs}(x)=\bar{\psi}_r(x)\gamma_5\gamma^0\psi_s(x),
\quad
P_{rs}(x)=\bar{\psi}_r(x)\gamma_5\psi_s(x)
\nonumber \\ \nonumber \\
&&V^\mu_{rs}(x)=\bar{\psi}_r(x)\gamma^\mu\psi_s(x),
\quad
T^{\mu\nu}_{rs}(x)=\bar{\psi}_r(x)\gamma^\mu\gamma^\nu\psi_s(x)
\nonumber \\ \nonumber \\ \nonumber \\
&&\mathcal{A}^0_{rs}(x)=A^0_{rs}(x)+ac_A\frac{\partial_0+\partial_0^*}{2}P_{rs}(x)
\nonumber \\ \nonumber \\
&&\mathcal{V}^\mu_{rs}(x)=V^\mu_{rs}(x)+ac_V\frac{\partial_\nu+\partial_\nu^*}{2}T^{\mu\nu}_{rs}(x)
\nonumber
\end{eqnarray}
where $r$ and $s$ are flavour indexes while $\zeta$ and $\zeta^\prime$ are boundary fields 
at $x_0=0$ and $x_0=T$ respectively.
The improvement coefficients $c_A$ and $c_V$ have been taken from 
refs.~\cite{Luscher:1996ug,Guagnelli:1997db,Sint:1997jx}.
We have calculated the following correlation functions
\begin{eqnarray}
&&\mathcal{F}_{i\rightarrow f}^\mu(x_0;{\bf p_i},{\bf p_f})= \frac{a^3}{2}\sum_{\bf x}{
\langle O_{li} \ \mathcal{V}^\mu_{if}(x)\ O_{fl}^\prime \rangle
}
\nonumber \\ \nonumber \\
&&f^{\mathcal{A}}_f(x_0,{\bf p_f})=-\sum_{{\bf x}}{
\langle O_{lf} \mathcal{A}^0_{fl}(x)\rangle
} \nonumber
\end{eqnarray}
where $i$ and $f$ refer to the heavy flavours while $l$ to the light one. 
The external momenta have been set by using flavour twisted b.c. for the 
heavy flavours; in particular we have used
\begin{eqnarray}
\psi_{i,f}(x+\hat{1}L)=e^{i\theta_{i,f}}\psi_{i,f}(x)
\nonumber
\end{eqnarray}
leading to
\begin{eqnarray}
p_1=\frac{\theta_{i,f}}{L}+\frac{2\pi k_1}{L},\qquad k_1\in \mathbb{N}
\nonumber
\end{eqnarray}
and ordinary periodic b.c. in the other spatial directions and for the light quarks. 
We work in the Lorentz frame in which the parent particle is at rest (${\bf p_i=0}$).
In this frame $w$ is obtained from the ratio between the energy and the
mass of the final particle $w=E_f/M_f$.
\begin{table}[t]
\begin{ruledtabular}
\begin{tabular}{cccc}
 & $\beta$ & $T\times L^3$ & $N_{cnfg}$\\
\hline
$L_0A$ & 7.300 & $48\times 24^3$   & 277  \\
$L_0B$ & 7.151 & $40\times 20^3$   & 224  \\
$L_0C$ & 6.963 & $32\times 16^3$   & 403  \\
\hline
$L_0a$ & 6.737 & $24\times 12^3$   & 608  \\
$L_0b$ & 6.420 & $16\times  8^3$   & 800  \\
$L_1A$ & 6.737 & $48\times 24^3$   & 260  \\
$L_1B$ & 6.420 & $32\times 16^3$   & 350  \\
\hline
$L_1a$ & 6.420 & $32\times 16^3$   & 360  \\
$L_1b$ & 5.960 & $16\times  8^3$   & 480  \\
$L_2A$ & 6.420 & $48\times 24^3$   & 250  \\
$L_2B$ & 5.960 & $24\times 12^3$   & 592  \\
\end{tabular}
\end{ruledtabular}
\caption{\label{tab:sims}
Table of lattice simulations.}
\end{table}

By assuming ground state dominance and by relying on the conservation of the vector current,
the matrix elements of $V^\mu$ can be extracted by considering the ratio
\begin{eqnarray}
\langle V^\mu \rangle_{D1}^{i\rightarrow f}\equiv
\langle \mathcal{M}_f\vert \ V^\mu \ \vert \mathcal{M}_i\rangle_{D1}=
\nonumber \\ \nonumber \\
2\sqrt{M_i E_f}\frac{\mathcal{F}_{i\rightarrow f}^\mu(T/2;{\bf 0},{\bf p_f})}{
\sqrt{
\mathcal{F}_{i\rightarrow i}^0(T/2;{\bf 0},{\bf 0})
\mathcal{F}_{f\rightarrow f}^0(T/2;{\bf p_f},{\bf p_f})
}}
\nonumber \\ \nonumber \\
\label{eq:def1}
\end{eqnarray}
An alternative definition of the matrix elements ($D2$), which 
reduces to the previous one ($D1$) in the limits of infinite volume and
zero lattice spacing, can be obtained by considering
\begin{eqnarray}
\langle V^\mu \rangle_{D2}^{i\rightarrow f}\equiv
\langle \mathcal{M}_f\vert \ V^\mu \ \vert \mathcal{M}_i\rangle_{D2}=
\nonumber \\ \nonumber \\
2\frac{\sqrt{M_i} E_f f_A^f(T/2,{\bf 0})}{\sqrt{M_f}f_A^f(T/2,{\bf p_f})}
\frac{\mathcal{F}_{i\rightarrow f}^\mu(T/2;{\bf 0},{\bf p_f})}{
\sqrt{
\mathcal{F}_{i\rightarrow i}^0(T/2;{\bf 0},{\bf 0})
\mathcal{F}_{f\rightarrow f}^0(T/2;{\bf 0},{\bf 0})
}}
\nonumber \\ \nonumber \\
\label{eq:def2}
\end{eqnarray}
In eqs.~(\ref{eq:def1}) and~(\ref{eq:def2}) the renormalization factors $Z_V$ and $Z_A$
cancel in the ratios together with factors containing the improvement coefficients
$b_V$ and $b_A$.

By introducing the following ratio
\begin{eqnarray}
x_f&=&\frac{\mathcal{F}_{f\rightarrow f}^1(T/2;{\bf 0},{\bf p_f})}
{\mathcal{F}_{f\rightarrow f}^0(T/2;{\bf 0},{\bf p_f})}
\nonumber \\ \nonumber \\
&=& \frac{\langle \mathcal{M}_f\vert \ V^1 \ \vert \mathcal{M}_f\rangle}
{\langle \mathcal{M}_f\vert \ V^0 \ \vert \mathcal{M}_f\rangle}
= \frac{\sqrt{w^2-1}}{w+1} \nonumber
\end{eqnarray}
we can define $w$, as well as  $E_f$ and $M_f$, entirely in terms of three point
correlation functions. This definition of $w$ is noisier than the one that can be obtained
in terms of ratios of two point correlation functions; however it leads to exact vector
current conservation when $M_f=M_i$ and reduces the final statistical error on the
form factors. The two definitions of the matrix elements lead to two definitions of 
$G^{i\rightarrow f}$
\begin{eqnarray}
G^{i\rightarrow f} &=& \frac{\sqrt{r}\langle V^0 \rangle^{i\rightarrow f}}{M_i(1+r)}\left\{
1\;
+ \;\frac{wr-1}{r\sqrt{w^2-1}}
\;
\frac{\langle V^1 \rangle^{i\rightarrow f}}
{\langle V^0 \rangle^{i\rightarrow f}}\right\}
\nonumber \\ \nonumber \\
r&=&\frac{M_f}{M_i}
\label{eq:fdefme}
\end{eqnarray}
The last equation is not defined at $w=1$; this is due
to the second term in parenthesis that we extrapolate at zero recoil before calculating 
$G^{i\rightarrow f}(w=1)$.

\section{the step scaling method}
\begin{figure}[t]
\includegraphics[width=0.48\textwidth]{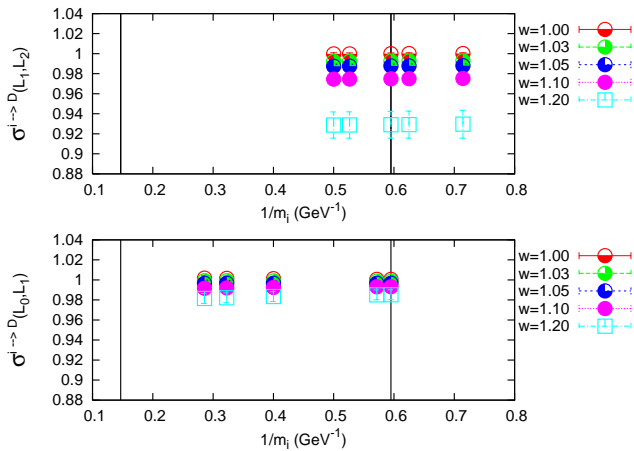}
\caption{\label{fig:sigmas} Step scaling functions $\sigma^{i\rightarrow D}(w;L_0,L_1)$,
lower plot, and $\sigma^{i\rightarrow D}(w;L_1,L_2)$, upper plot, as functions of
the inverse of the RGI heavy quark mass $m_i$ of the parent meson measured in GeV. The black 
vertical lines represent the physical values of $m_c$ and $m_b$.
The data correspond to the definition $D1$.}
\end{figure}
The SSM has been introduced to cope with two-scale problems in lattice QCD. In the
calculation of heavy-light meson properties the two scales are the mass
of the heavy quarks ($b$,$c$) and the mass of the light quarks ($u$,$d$,$s$). 
Here we consider the form factor $G^{i\rightarrow f}$ as a 
function of $w$, the volume, $L^3$, and identify heavy meson states by the corresponding RGI 
quark masses that in the infinite volume limit lead to the physical meson 
spectrum~\cite{deDivitiis:2003iy};  
the RGI quark masses are measured by the lattice version of the PCAC relation and are not
affected by finite volume effects (see~\cite{comingsoon} for further details). 

First we compute the observable $G^{B\rightarrow D}(w;L_0)$ on a small volume, $L_0$, 
which is chosen to accommodate the dynamics of the $b$-quark. As in our previous work we fixed
$L_0=0.4$~fm.
Then we evaluate a first effect of finite volume by evolving the volume from $L_0$ to 
$L_1=0.8$~fm and computing the ratio
\begin{displaymath} 
\sigma^{i\rightarrow D}(w;L_0,L_1)=\frac{G^{i\rightarrow D}(w;L_1)}{G^{i\rightarrow D}(w;L_0)}
\end{displaymath}
The crucial point is that the step scaling functions are calculated by simulating
heavy quark masses $m_i$ smaller than the $b$-quark mass. The physical
value $\sigma^{B\rightarrow D}(w;L)$ is obtained by a smooth extrapolation in $1/m_i$ that
relies on the HQET expectations and upon the general idea that finite volume effects, measured
by the $\sigma$'s, are almost insensitive to the high energy scale. 
The final result is obtained by further evolving the volume from $L_1$ to $L_2=1.2$~fm,
according to
\begin{eqnarray} 
&&G^{B\rightarrow D}(w;L_2)=
\nonumber \\ \nonumber \\
&&=G^{B\rightarrow D}(w;L_0)\
\sigma^{B\rightarrow D}(w;L_0,L_1)\
\sigma^{B\rightarrow D}(w;L_1,L_2)
\nonumber
\end{eqnarray}
Physical values require also usual continuum and chiral extrapolations.

\begin{figure}[t]
\includegraphics[width=0.48\textwidth]{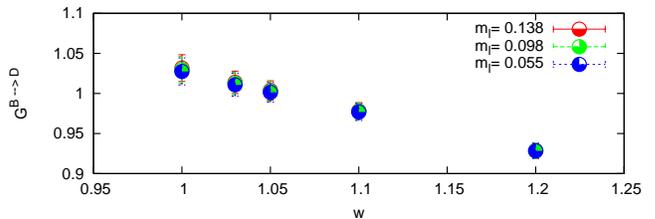}
\caption{\label{fig:chiral} Light quark mass dependence of $G^{B\rightarrow D}(w;L_0)$
in the continuum limit.
The different sets of points correspond to different values of $m_l$ ranging from
about $m_s$ to about $m_s/4$.
The data correspond to the definition $D1$.} 
\end{figure}
\begin{figure}[t]
\includegraphics[width=0.48\textwidth]{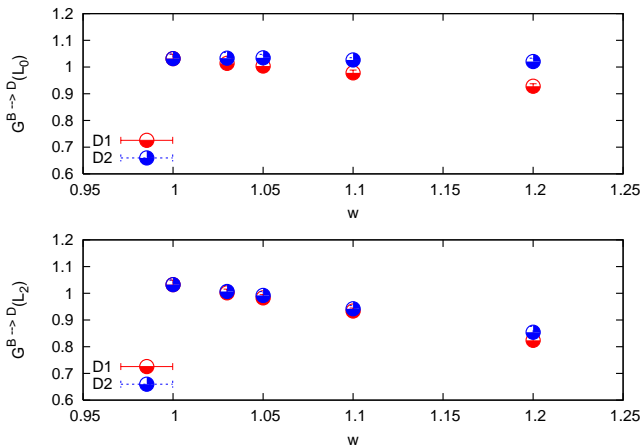}
\caption{\label{fig:2defs} Comparison of the two definitions of $G^{B\rightarrow D}(w;L)$
at $L_0=0.4$~fm (upper plot) and at $L_2=1.2$~fm (lower plot).} 
\end{figure}
%

\section{results}
In figure~\ref{fig:sigmas} we can test the validity of the SSM. The step scaling functions
are almost insensitive to the heavy quark mass $m_i$ of the parent meson for values larger 
than $m_c$. Moreover, finite volume effects are already very small at $L=0.8$~fm, in 
particular at zero recoil. 

We present results already extrapolated to the chiral and continuum limits.
We have simulated three different lattices for $L_0$ and two for the other volumes 
(see table~\ref{tab:sims}) observing small $O(a)$ effects on the form factor 
and on the step scaling functions. 
Further details on the continuum extrapolations will be given in ref.~\cite{comingsoon}.
Concerning the chiral behaviour, our results do not show any
sizable dependence upon the light quark mass, as shown in figure~\ref{fig:chiral}
for $G^{B\rightarrow D}(w;L_0)$. The same feature is observed for all the combinations of
heavy quark masses and on all the volumes that we have simulated.
Nevertheless we make a linear extrapolation to reach the chiral limit;
the resulting error largely accounts for the systematics due to these extrapolations.

As discussed in sec.~\ref{sec:notations}, we used two definitions of the form factor
that, at finite volume, differ by the finite volume effects. A check of the convergence of
our SSM can be obtained by comparing these two definitions on the smallest volume 
(figure~\ref{fig:2defs} upper plot) and on the largest one (figure~\ref{fig:2defs} 
lower plot).
We see that, while the small volume results differ, the final ones converge to a common
value giving us confidence  of a correct accounting of finite volume effects. 

\begin{table}[t]
\begin{ruledtabular}
\begin{tabular}{cccc}
$w$ & $G^{B\rightarrow D}(w)$ & $N_f$ & reference\\
\hline
\\[-5pt]
1.00 & 1.026(17) & 0   & this work \\
1.03 & 1.001(19) & 0   & this work \\
1.05 & 0.987(15) & 0   & this work \\
1.10 & 0.943(11) & 0   & this work \\
1.20 & 0.853(21) & 0   & this work \\
\\
\hline
1.00 & 1.058(20) & 0   & \cite{Hashimoto:1999yp} \\
1.00 & 1.074(24) & 2+1 & \cite{Okamoto:2004xg}   \\
\end{tabular}
\end{ruledtabular}
\caption{\label{tab:gw}
Final result. Form factor $G^{B\rightarrow D}(w)$ in the continuum and
infinite volume limits. As a comparison we quote also the results of
previous lattice calculations by the Fermilab collaboration.}
\end{table}
%

\section{concluding remarks}
In table~\ref{tab:gw} we quote our final results obtained by averaging over the two definitions and by combining in quadrature statistical errors with the systematic ones due to the small residual dispersion between $D1$ and $D2$. 
As a comparison we show previous lattice results obtained by the Fermilab 
collaboration at zero recoil. 

\begin{figure}[t]
\includegraphics[width=0.48\textwidth]{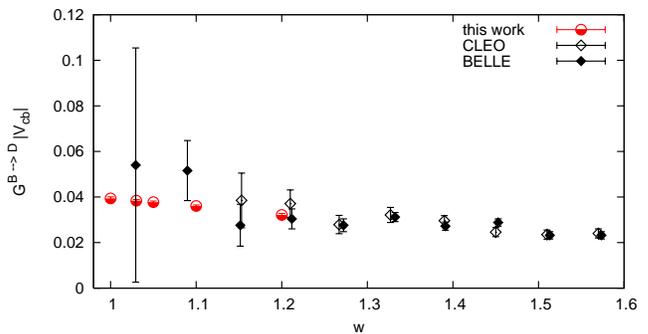}
\caption{\label{fig:vcb} Comparison of $|V_{cb}|\ G^{B\rightarrow D}(w)$ 
with available experimental data. The plot has been done by normalizing our lattice data 
with the value of $V_{cb}$ extracted at $w=1.2$.} 
\end{figure}

The existence of predictions up 
to $w\simeq 1.2$ and physical $b$ and $c$ quark masses allows a direct comparison with 
experimental data, as shown in figure~\ref{fig:vcb}. 
The comparison has been done by extracting the value of $V_{cb}$ by the ratio of the 
experimental and lattice data at $w=1.2$; as an indication, we get 
$V_{cb}=3.84(9)(42)\times10^{-2}$,
where the first error is from our lattice result, $G^{B\rightarrow D}(w=1.2)=0.853(21)$,
and the second from the experimental decay rate, 
$|V_{cb}|G^{B\rightarrow D}(w=1.2)=0.0327(35)$, as deduced from the plots of 
refs.~\cite{Belle,CLEO,Albertus:2005vd}.

The extension of this calculation to the unquenched case does not present problems of principle. The recursive matching process can be
extended to the sea quark masses that, alternatively, can be kept to their physical values
if the Schr\"odinger Functional formalism is used.
Moreover, flavour twisted boundary conditions can be used for heavy valence quarks 
also in the $N_f=3$ unquenched theory. The real case will further differ by the
heavy flavour determinants that can be computed perturbatively.
%

\begin{acknowledgments}
The simulations required to carry on this project 
have been performed on the INFN apeNEXT machines at Rome "La Sapienza".
We warmly thank A.~Lonardo, D.~Rossetti and P.~Vicini for technical advice. 
\end{acknowledgments}


\end{document}